\documentclass[12pt]{article}

\usepackage{amsmath}
\usepackage{graphicx}

\usepackage{natbib}
\bibpunct[,]{(}{)}{;}{a}{}{,}

\newcommand{\expect}[1]{\ensuremath{\mathrm{E}[#1]}}

\begin{document}

\title{Population genetics of translational robustness}
\author{Claus O. Wilke$^*$ and D. Allan Drummond$^\dagger$\\\mbox{}\\$^*$Section of Integrative Biology\\ and Center for Computational Biology and Bioinformatics,\\University of Texas, Austin, TX 78712, USA\\$^\dagger$Program in Computation and Neural Systems,\\California Institute of Technology,\\ Pasadena, CA 91125, USA}

\maketitle

\newpage
\noindent
Running head: Translational robustness
\bigskip

\noindent
Keywords: translation, evolutionary rate, expression level, neutrality, protein evolution

\bigskip

\noindent
Corresponding author:\\
\mbox{}~~Claus O. Wilke\\
\mbox{}~~Integrative Biology\\
\mbox{}~~\#1 University Station -- C0930\\
\mbox{}~~University of Texas, Austin, TX 78712, USA\\
\mbox{}~~cwilke@mail.utexas.edu\\
\mbox{}~~Phone: (512) 471 6028\\
\mbox{}~~Fax: (512) 471 3878

\bigskip

\noindent
\textbf{Abstract:} Recent work has shown that expression level is the main predictor of a gene's evolutionary rate, and that more highly expressed genes evolve slower. A possible explanation for this observation is selection for proteins which fold properly despite mistranslation, in short selection for translational robustness. Translational robustness leads to the somewhat paradoxical prediction that highly expressed genes are extremely tolerant to missense substitutions but nevertheless evolve very slowly. Here, we study a simple theoretical model of translational robustness that allows us to gain analytic insight into how this paradoxical behavior arises.

\newpage

\centerline{INTRODUCTION}
\bigskip

The increasing availability of whole-genome sequences from many different species has revealed a surprising fact: Different genes within the same organism evolve at dramatically different rates. For example, the evolutionary rates of the fastest- and slowest-evolving genes in \emph{Saccharomyces cerevisiae} are separated by three orders of magnitude \citep{Drummondetal2005}. Because the dominant force shaping genome-wide patterns of evolutionary rate is most likely stabilizing selection, the evolutionary rates of genes should correlate with quantities that measure how important or otherwise constrained a gene is \citep{Kimura83,Ohta92}. A wide array of such quantities have been proposed, shown to correlate with evolutionary rate, and subsequently disputed. Examples include a gene's dispensability or essentiality \citep{HurstSmith99,HirshFraser2001,Jordanetal2002,Paletal2003,ZhangHe2005,Walletal2005}, its number of interaction partners \citep{Fraseretal2002,BloomAdami2003,Jordanetal2003,Hahnetal2004,Agrafiotietal2005}, its length \citep{MaraisDuret2001}, or its centrality in the protein interaction network \citep{HahnKern2005}. However, it seems that most importantly, the expression level \citep{Paletal2001,RochaDanchin2004}, or perhaps more accurately the frequency of translation events \citep{Drummondetal2005,Drummondetal2005b}, influence evolutionary rate.

\citet{Drummondetal2005} have recently introduced a theory for why highly expressed genes evolve slowly. Translation is error prone, and inactivated or misfolded proteins resulting from mistranslation impose substantial costs on the cell \citep{Bucciantinietal2002}, costs which increase with expression level.  One way in which the cost associated with a highly expressed gene is reduced is translational accuracy \citep{Akashi94,Akashi2001}, whereby the gene is encoded with optimal codons whose translation is less error-prone than the translation of other codons. Translational accuracy can explain why the rate of synonymous evolution $dS$ is correlated with expression level, codon adaptation index, or protein abundance. However, it cannot explain why the rate of non-synonymous evolution $dN$ shows an even stronger correlation with these quantities \citep{Drummondetal2005}. Selection for translational accuracy can reduce the translational error rate by a factor of 4--9 \citep{PrecupParker87}, but even optimally coded genes that are highly expressed may produce a large amount of erroneous polypeptides. Therefore, \citet{Drummondetal2005} suggest that highly expressed genes should be under additional selection to be tolerant to translation errors, that is, the polypeptides produced from these genes should fold properly even if they were erroneously translated. Recent work on protein biochemistry has shown that proteins can differ widely in their tolerance to missense substitutions, and that properly chosen point mutations can dramatically increase the tolerance of a protein to additional substitutions \citep{Bloometal2005}. \citeauthor{Drummondetal2005}'s hypothesis, referred to as selection for translational robustness, predicts a constraint on the non-synonymous rate of evolution, whereas selection for translational accuracy predicts primarily a constraint on the synonymous rate of evolution. We must assume that both selective constraints will operate on genes that are frequently translated.

Genes that are highly tolerant to translational missense errors must also, by definition, be highly tolerant to missense substitutions. However, the translational robustness hypothesis predicts that these genes will nevertheless be strongly conserved under evolution. An example of a gene that exhibits this paradoxical behavior is Rubisco, an extremely abundant protein which fixes carbon dioxide during photosynthesis. Rubisco is strongly conserved across phyla, but appears to tolerate most missense substitutions in the laboratory without loss of fold \citep{Spreitzer93,KelloggJuliano1997}.

The purpose of the present paper is to put the translational robustness hypothesis into precise mathematical terms, and to demonstrate how highly expressed genes can evolve to be tolerant to missense substitutions and yet remain strongly conserved under evolution.

\bigskip

\centerline{MATERIALS AND METHODS}
\bigskip

\noindent
\textbf{Model:} We consider the evolution of a gene encoding a protein of length $L$. Each site in the protein can be in one of two states, \emph{neutral} or \emph{non-neutral}. We denote the number of neutral sites in the protein by $n$. A mutation at a neutral site of a folded protein leaves the protein folded, but changes the site from neutral to non-neutral. A mutation at a non-neutral site of a folded protein will usually cause misfolding and consequent loss of function, but with a small probability $\alpha$, such a mutation will not affect folding but turn the site into a neutral one. For simplicity, we assume that once an amino-acid sequence loses the ability to fold, it is impossible to mutate it back into a folded state. The rationale behind this assumption is that the likelihood of a mutation restoring fold to an unfolded amino-acid sequence is so low as to be negligible. Our model is a reasonable abstraction of a thermodynamic framework of protein evolution that has recently been shown to have good predictive power for both simulated and real proteins \citep{Bloometal2005,Wilkeetal2005}. The key insight of this framework is that a protein's tolerance to substitutions is closely related to the protein's stability---more stable proteins can withstand more missense substitutions---and that therefore proteins can change from being highly fragile to being highly robust to mutations and vice versa through the accumulation of stabilizing or destabilizing mutations. In this sense, a mutation from a non-neutral to a neutral site in our model corresponds to a stabilizing mutation, and the opposite mutation corresponds to a destabilizing mutation. Thus, our model captures the following key aspects of protein biochemistry: (i) Homologous proteins can vary widely in their tolerance to mutations, and individual point mutations can increase or decrease this tolerance. (ii) Mutations that increase a protein's tolerance are much rarer than mutations that decrease its tolerance. (iii) Highly tolerant proteins are extremely rare, while moderately tolerant proteins are abundant.  (iv) Non-functional mutant proteins are likely to be misfolded.

The gene is expressed at a level that leads to the synthesis of $x$ polypeptides. For simplicity, we assume that the total number of polypeptides translated per gene is proportional to the gene's expression level, and that the constant of proportionality is 1. Thus, $x$ is also the expression level, measured in mRNA molecules per cell. Under translation, each site is mistranslated with probability $\tau$ (we neglect premature termination of the translation process). The probability that a single mRNA molecule is mistranslated and leads to a misfolded protein is $1-[1-\tau(1-\alpha)]^{L-n}\approx\tau(L-n)$, where the approximation holds for $\alpha,\tau\ll1$. Let $f=\tau(L-n)$ be the fraction of synthesized proteins that misfold. We assume that the expression level is regulated such that the number of folded proteins per gene, the protein abundance $A$, is held constant, regardless of the translational error rate. Then, $A=x(1-f)$. The total number of misfolded polypeptides per gene follows as $xf=Af/(1-f)$. Finally, we assume that each misfolded polypeptide imposes a cost $c$ on the cell, so that the total cost of a gene translated at abundance level $A$ is $cAf/(1-f)$. We turn this cost into a fitness value by assuming that each misfolded protein has the same relative effect on fitness. Then we can write the overall fitness of a gene with $n$ neutral sites as
\begin{equation}\label{eq:fitness-def}
  w_n = \exp\Big(-cA\frac{\tau(L-n)}{1-\tau(L-n)}\Big)\,.
\end{equation}
Without loss of generality, we use $c=1$ henceforth.

\bigskip

\noindent
\textbf{Simulations:} We implemented a stochastic simulation of $N$ sequences reproducing in discrete, non-overlapping generations. We employed standard Wright-Fisher sampling, that is, the probability that a sequence in generation $t+1$ is the offspring of a sequence at generation $t$ is directly proportional to the latter's fitness.

We measured the evolutionary rate along the line of descent from the most recent common ancestor (MRCA) of the final population backwards in time, as described by \citet{Wilke2004}.  Briefly, we let the simulated population evolve until the birth-time of the population's MRCA, designated $t_0$, exceeded a fixed equilibration time $t_\text{equil}$ plus a time window $t_{\text{meas}}$, $t_0>t_\text{equil}+t_{\text{meas}}$.  All quantities were measured on the equilibrated population during this latter time window.  For all results reported, we chose $t_\text{equil}=t_\text{meas}=400000$, $U=0.001$, $\tau=0.001$, and $L=100$. At all parameter settings, we carried out 100 replicas and averaged results over all replicas.

\bigskip

\centerline{ANALYTICAL RESULTS}
\bigskip

\noindent
\textbf{Solution based on Sella-Hirsh theory:} We can calculate the steady-state solution of our model using the analogy between evolutionary biology and statistical physics recently demonstrated by \citet{SellaHirsh2005}. The theory of \citet{SellaHirsh2005} is applicable whenever the product of population size and mutation rate is much smaller than one, $NU\ll 1$. In this regime, the population is essentially homogeneous at all times, and can be represented at any given point in time by a single sequence. We say the population is in state $i$ if the dominant sequence in the population is sequence $i$. The key insight of \citet{SellaHirsh2005} is that the probability $p_i$ to find the population in state $i$ is proportional to a function $F(i)$ (also called a Boltzmann factor) that depends only on the fitness of sequence $i$, the population size, and details of the mutation process. Thus, it follows that
\begin{equation}\label{eq:sella-hirsh}
  p_i = F(i)/\sum_j F(j)\,, 
\end{equation}
where the sum runs over all possible sequences $j$. Once we have the probabilities $p_i$, we can calculate all observable quantities of interest, such as expected fitness and expected evolutionary rate, using standard probability theory (see also below).

As the fitness of a sequence in our model depends only on the sequence's number of neutral sites $n$, it is useful to lump all sequences with the same $n$ into a single class, and calculate the probability $p_n$ that the population is in a state represented by any sequence with $n$ neutral sites. Since there are $\binom{L}{n}$ such sequences, we introduce $F'(n)=\binom{L}{n}F(n)$, where $F(n)$ is the appropriate Boltzmann factor for a single sequence with $n$ neutral sites, and then calculate $p_n$ as $p_n=F'(n)/\sum_{k=0}^L F'(k)$. In our case, $F'(n)$ is given by
\begin{equation}\label{eq:Boltzmann-factor}
F'(n) =\binom{L}{n}\exp[(2N-2)\ln(w_n)+n\ln(\alpha)]
\end{equation}
with $w_n$ as defined in Eq.~\eqref{eq:fitness-def}. The second term in the exponential takes into account the asymmetry in the mutation process, that is, mutations that increase $n$ are a factor $\alpha$ less likely to occur than mutations that decrease $n$ (\citealt{SellaHirsh2005}, supplementary text).

With the formalism outlined in the previous paragraphs, we can calculate the expected number of neutral sites in the steady state $\expect{n}$ as
\begin{equation}\label{eq:expect-n}
  \expect{n} = \sum_{n=0}^L n F'(n)/\sum_{k=0}^L F'(k)\,,
\end{equation}
and the expected fitness as
\begin{equation}\label{eq:expect-w}
\expect{w} = \sum_{n=0}^L w_n F'(n)/\sum_{k=0}^L F'(k)\,.
\end{equation}
Note that the expected values are not taken over the population (which is assumed to be homogeneous), but over a long time-window in the steady state. Next we can calculate the evolutionary rate $K$, that is, the expected number of amino-acid substitutions per unit time that accumulate along the line of descent in an equilibrated population. We find
\begin{equation}\label{eq:evol-rate}
K = \sum_{n=0}^L NU[\alpha\pi(n\rightarrow n+1)+\pi(n\rightarrow
n-1)]F'(n)/\sum_{k=0}^L F'(k)\,,
\end{equation}
where $\pi(i\rightarrow j)$ is the probability of fixation of sequence $j$ in background $i$ \citep{SellaHirsh2005},
\begin{equation}\label{eq:prob-fix}
  \pi(i\rightarrow j) = \frac{1-e^{-2\ln(w_j/w_i)}}{1-e^{-2N\ln(w_j/w_i)}}\,.
\end{equation}
[Note that $\pi( L \rightarrow L+1) := 0$ and $\pi( 0 \rightarrow -1) := 0$.]

We can simplify these expressions in the special cases that $A$ is either very large or very small. After inserting Eq.~\eqref{eq:fitness-def} into Eq.~\eqref{eq:Boltzmann-factor}, we have
\begin{equation}\label{eq:Boltzmann-explicit}
  F'(n) = \binom{L}{n}\exp\Big[-2NA\frac{\tau(L-n)}{1-\tau(L-n)}+n\ln(\alpha)\Big]\,,
\end{equation}
where we have also made the approximation $N-1\approx N$. We will continue to use this approximation throughout the rest of the paper. From Eq.~\eqref{eq:Boltzmann-explicit}, we can see that the behavior of the system changes drastically depending on whether the product $NA$ is large or small. However, since the population size $N$ is the same for all genes in a species while each gene's corresponding protein abundance $A$ can vary over many orders of magnitude, in the following we assume that $N$ is fixed and consider the limits of very small and very large $A$.

For $A\rightarrow0$, the first term in the exponential disappears, and $F'(n)$ becomes simply $\binom{L}{n}\alpha^n$. Thus, we find
\begin{align}\label{eq:nave-smallA}
  \expect{n} &= \sum_{n=0}^L n \binom{L}{n} \alpha^n / \sum_{n=0}^L \binom{L}{n} \alpha^n\notag\\
  &= \alpha L/(\alpha+1)\,.
\end{align}
We cannot obtain similarly simple expressions for $\expect{w}$ and $K$ in this limit, but will do so in the next subsection using a different method.

For $A\rightarrow\infty$, we have to distinguish between the cases $n=L$ and $n<L$. For $n<L$, the first term in the exponential in Eq.~\eqref{eq:Boltzmann-explicit} becomes much larger in magnitude than the second term, which is a constant. Thus, we can neglect the second term, and have
\begin{equation}
  F'(n)=\binom{L}{n}\exp\Big[-2NA\frac{\tau(L-n)}{1-\tau(L-n)}\Big]\,.
\end{equation}
This expression tends to 0 for large $A$. For $n=L$, we have $F'(L)=\alpha^L$, independent of $A$. All terms but the $n=L$ term disappear, and we have $\expect{n}=L$, $\expect{w}=w_L$, and
\begin{align}\label{eq:Kave-largeA}
  K &=NU\pi(L\rightarrow L-1)\notag\\
  &=NU e^{-2NA\tau/(1-\tau)}\,.
\end{align}

\bigskip

\noindent
\textbf{Approximate solution:} Sella-Hirsh theory yields an exact solution for our model. However, the resulting expressions are somewhat unwieldy, and don't lead to simple analytical expressions for intermediate $A$. Therefore, we will now derive an approximate solution to our model.

For small $\tau$, we can approximate $w_i \approx \exp[-A\tau(L-i)]$, and find $\ln(w_j/w_i) = A\tau(j-i)$. This approximation is equivalent to neglecting the small number of additional translation events required to replace polypeptides that misfold. The probability of fixation follows from Eq.~\eqref{eq:prob-fix} as
\begin{equation}\label{eq:pi-approx}
  \pi(i\rightarrow j) = \frac{1-e^{-2A\tau(j-i)}}{1-e^{-2NA\tau(j-i)}}\,.
\end{equation}
The idea of the approximate solution is that in the steady state, mutations that increase the number of neutral sites and those that decrease it are in perfect balance. Therefore, the number of neutral sites in the steady state, $n^*$, satisfies:
\begin{equation}\label{eq:equil}
  \alpha(L-n^*)\pi(n^*\rightarrow n^*+1) = n^*\pi(n^*\rightarrow n^*-1)\,.
\end{equation}
According to Eq.~\eqref{eq:pi-approx}, $\pi(n\rightarrow n+1)$ and $\pi(n\rightarrow n-1)$ are independent of $n$. We introduce $\pi_+$ and $\pi_-$, the probabilities of fixation for a mutation that increases or decreases $n$ by one, respectively, and find:
\begin{align}
  \pi_+ &=\frac{1-e^{-2A\tau}}{1-e^{-2NA\tau}}\,,\\
  \pi_- &=\frac{1-e^{2A\tau}}{1-e^{2NA\tau}}\,.
\end{align}
After inserting these expressions into Eq.~\ref{eq:equil} and solving for $n^*$, we obtain
\begin{equation}\label{eq:n-approx}
  n^* = \frac{\alpha L \pi_+}{\alpha \pi_+ + \pi_-}\,.
\end{equation}
With this result, the expected fitness in the steady state becomes
\begin{equation}\label{eq:ave-fitness-approx}
  \expect{w} = \exp[-A\tau(L-n^*)]\,,
\end{equation}
and the evolutionary rate $K$ follows as
\begin{equation}\label{eq:approx-evol-rate}
  K=NU\Big(\alpha\pi_+\frac{L-n^*}{L} + \pi_-\frac{n^*}{L}\Big)\,.
\end{equation}
In the Appendix, we derive an expression for $K$ as a function of $n^*$, rather than as a function of $A$ as we have done here.

In the limit $A\rightarrow0$, we have $\pi_+=\pi_-= 1/N$ and $n^*= \alpha L/(\alpha+1)$\,. [Note that this expression is identical to the result found through Sella-Hirsh theory, Eq.~\eqref{eq:nave-smallA}, if we equate $n^*$ with $\expect{n}$.] Therefore, in this limit,
\begin{equation}\label{eq:Asmall}
  K = \alpha U\,.
\end{equation}
In the limit $A\rightarrow\infty$, we have $\pi_+= 1$, $\pi_-= e^{-2NA\tau}$, and $n^*= L$\,. Therefore, in this limit,
\begin{equation}\label{eq:Alarge}
  K = NU e ^{-2NA\tau}\,.
\end{equation}
The expressions for $n^*$ and $K$ again agree with the results found through Sella-Hirsh theory, if we assume $1-\tau\approx1$ in Eq.~\eqref{eq:Kave-largeA}.

\bigskip

\noindent
\textbf{Limitations on the number of neutral sites:} Certain residues may never tolerate any substitutions, such as the active-site serine of a serine protease or the heme-binding histidines of hemoglobin.  Under the assumption that $m$ sites can never be neutral, we can write $L=l+m$, and for small $\tau$ the fitness $w_n$ is approximately 
\begin{equation}\label{eq:m-rescale}
  w_n = e^{-A\tau(l+m-n)} = e^{-A\tau m}e^{-A\tau(l-n)}.
\end{equation}
In other words, all fitness values are rescaled by a factor depending on $\tau$ but not on $n$. Eq.~\eqref{eq:prob-fix} reveals that such a rescaling leaves the fixation probabilities unchanged. Therefore, the approximate solution remains unchanged except that we replace $L$ by $l$ ($=L-m$) everywhere and add a factor $e^{-A\tau m}$ to Eq.~\eqref{eq:ave-fitness-approx}.

For Sella-Hirsh theory, if we assume that $\tau m$ is negligibly small, the Boltzmann factor $F'(n)$ gains a similar leading factor which, as Eqs.~\eqref{eq:expect-n} and \eqref{eq:expect-w} make clear, also drops out, this time through the normalization term.  In the case of $\expect{n}$, the result is only that $L$ must again be replaced by $l=L-m$, while the expected value $\expect{w}$ also gains a leading factor $e^{-A\tau m}$, just as in the approximate case.

In short, when there are $m$ never-neutral sites, the main effects are to lower the population's fitness by a factor $e^{-A\tau m}$ and to reduce the expected number of neutral sites $\expect{n}$ and the evolutionary rate $K$ roughly as though the evolving gene were shorter by $m$ residues.

\bigskip

\centerline{SIMULATION RESULTS}
\bigskip

\begin{figure}
  \centerline{\includegraphics[width=4in,angle=270]{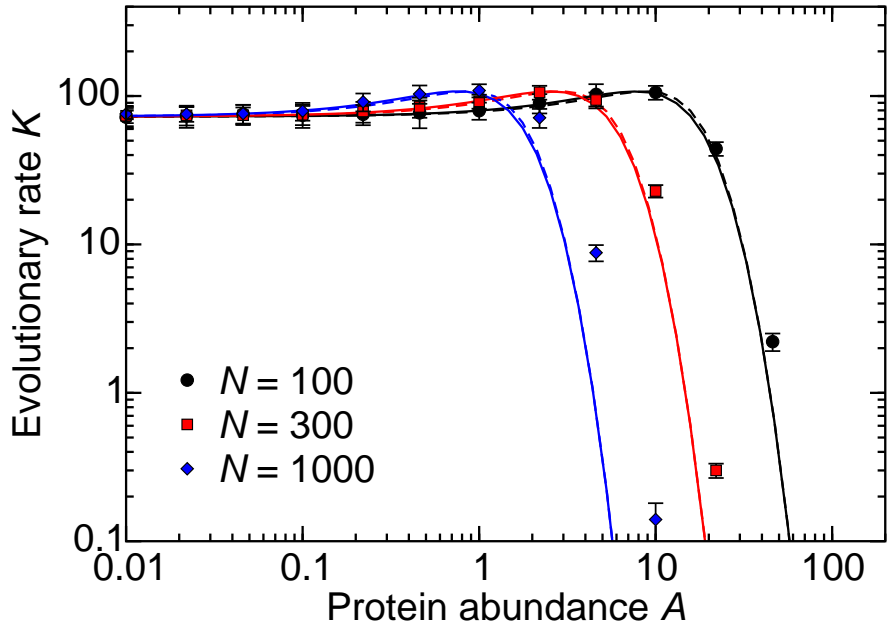}}
\caption{\label{fig:Nvar}Evolutionary rate $K$ (measured in substitutions per $4\times 10^5$ generations) versus the protein abundance $A$, for population sizes $N=100, 300, 1000$ ($\alpha=0.1$). Data points indicate simulation results, solid lines indicate the prediction using Sella-Hirsh theory, Eq.~\eqref{eq:evol-rate}, and dashed lines indicate the prediction using our approximate solution, Eq.~\eqref{eq:approx-evol-rate}. Error bars indicate standard errors.
}
\end{figure}

First, we studied the rate of evolution $K$ as a function of the protein abundance $A$, for various population sizes (Fig.~\ref{fig:Nvar}). We found that $K$ levels off for $A\rightarrow 0$. The asymptotic value at $A=0$ is $K(0)=\alpha U$ (Eq.~\ref{eq:Asmall}). For increasing $A$, $K$ first increases, and then rapidly drops off to zero for even larger $A$. The main effect of the population size $N$ is to determine at what level of $A$ this drop-off initiates. With increasing $N$, the evolutionary rate $K$ seems to be simply shifted to the left, towards lower $A$ (Fig.~\ref{fig:Nvar}). We can make this statement more precise by considering the large-$A$ limit of our approximate solution, Eq.~\ref{eq:Alarge}. This limit shows that the evolutionary rates $K(N,A)$ and $K(aN,a^{-1}A)$ are related through $K(N,A)\approx a^{-1}K(aN,a^{-1}A)$, where $a$ is an arbitrary constant. Therefore, if we increase $N$ by a factor of $a$, the resulting curve $K(A)$ appears on a log-log plot to be shifted upwards and to the left by an amount of $\log(a)$. The upwards shift cannot be noticed, because it exists only for very large $A$, and thus the effect of increasing $N$ seems to be to simply shift the $K(A)$ curve to the left.

\begin{figure}
  \centerline{\includegraphics[width=4in,angle=270]{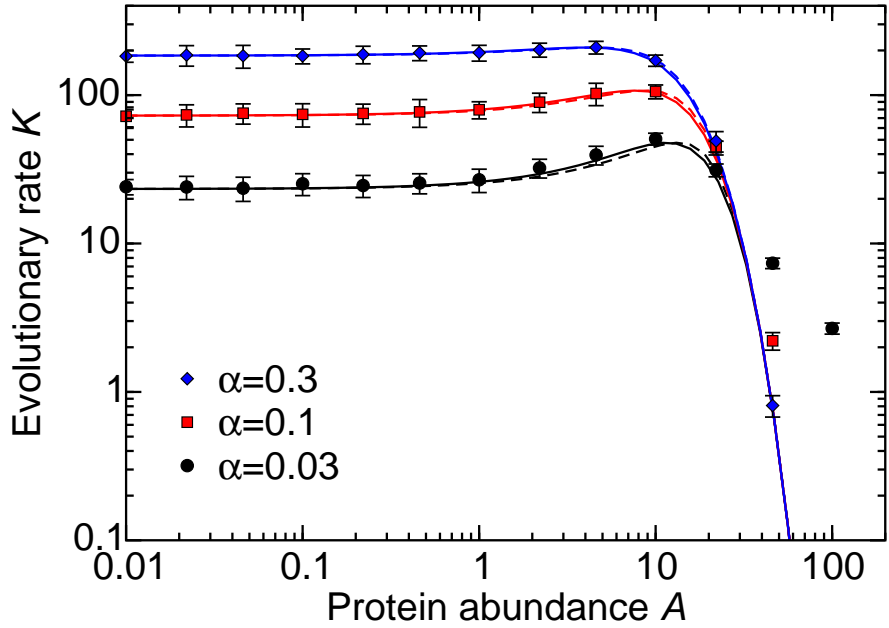}}
\caption{\label{fig:alphavar}Evolutionary rate $K$ (measured in substitutions per $4\times 10^5$ generations)  versus the protein abundance $A$, for values of $\alpha=0.03, 0.1, 0.3$ ($N=100$). Data points indicate simulation results, solid lines indicate the prediction using Sella-Hirsh theory, Eq.~\eqref{eq:evol-rate}, and dashed lines indicate the prediction using our approximate solution, Eq.~\eqref{eq:approx-evol-rate}. Error bars indicate standard errors. The deviation from the prediction of the two rightmost data points for $\alpha=0.03$ is caused by insufficient equilibration time.
}
\end{figure}

Second, we studied the effect of varying $\alpha$ on $K(A)$ (Fig.~\ref{fig:alphavar}). The variable $\alpha$ mainly influences the asymptotic limit of $K(A)$ for small $A$ (with $K(A)$ increasing with $\alpha$), but does not affect how quickly $K(A)$ decays for large $A$.

\begin{figure}
  \centerline{\includegraphics[width=4in,angle=270]{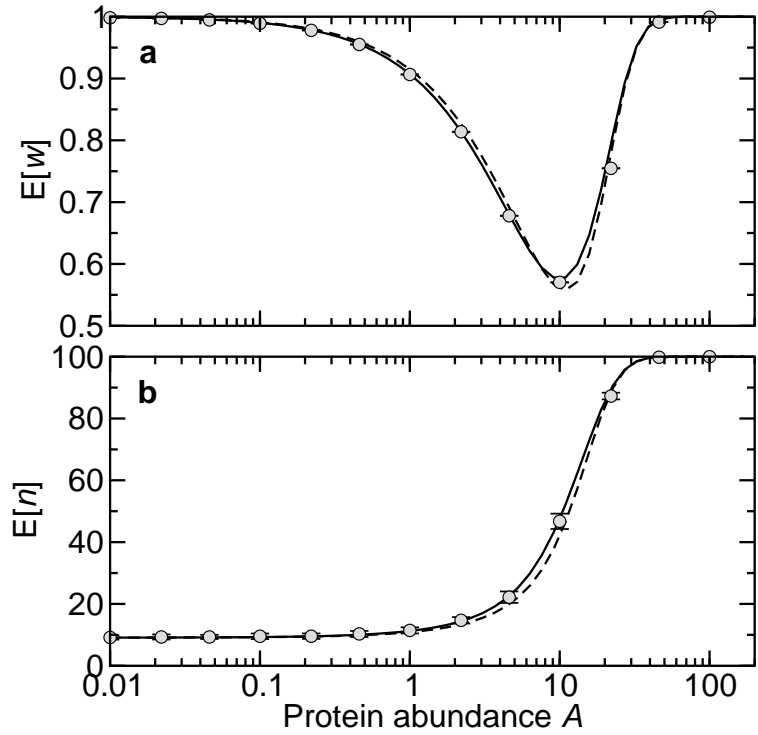}}
\caption{\label{fig:fitness-neutr}Expected fitness $\expect{w}$ (a) and expected number of neutral sites $\expect{n}$ (b) versus the protein abundance $A$, for $N=100$ and $\alpha=0.1$. Data points indicate simulation results, solid lines indicate the prediction using Sella-Hirsh theory, and dashed lines indicate the prediction using our approximate solution. Error bars indicate standard errors.
}
\end{figure}

Third, we studied the behavior of the expected fitness and the expected number of neutral sites for varying levels of $A$. The expected fitness is approximately 1 for both very low and very high $A$, but drops below 1 for the intermediate values of $A$ for which $K(A)$ starts to decay (Fig.~\ref{fig:fitness-neutr}a). The expected number of neutral sites is $\alpha L/(\alpha+1)$ for very small $A$, and increases to 1 for large $A$ (Fig.~\ref{fig:fitness-neutr}b). We can understand the different evolutionary regimes of low $A$ and high $A$ as follows: For low $A$, since very little protein is synthesized, the cost associated with misfolded proteins following erroneous translation is negligible. Therefore, the expected fitness in this regime is 1. Since the cost of misfolding is negligible, the number of neutral sites is not under selection in this regime, and it settles to the value at which the mutations increasing $n$ and those decreasing $n$ exactly balance each other. For high $A$, on the other hand, the cost of translation-induced misfolding is tremendous.  Therefore, at high $A$, the population converges to the single optimal sequence with $n=L$ (or $n=l$ if some sites can never be neutral). Every mutation that reduces $n$ by even 1 is highly deleterious, and therefore will virtually never go to fixation, even in a small population. For $l=L$, the optimal sequence (which has $n=l$) pays no cost whatsoever under mistranslation, and the expected fitness is again $1$.  For intermediate $A$, the cost of translation-induced misfolding is significant but not debilitating. As a result, $n$ is elevated in comparison to its low-$A$ limit, but the expected fitness falls below $1$.

\begin{figure}
  \centerline{\includegraphics[width=4in,angle=270]{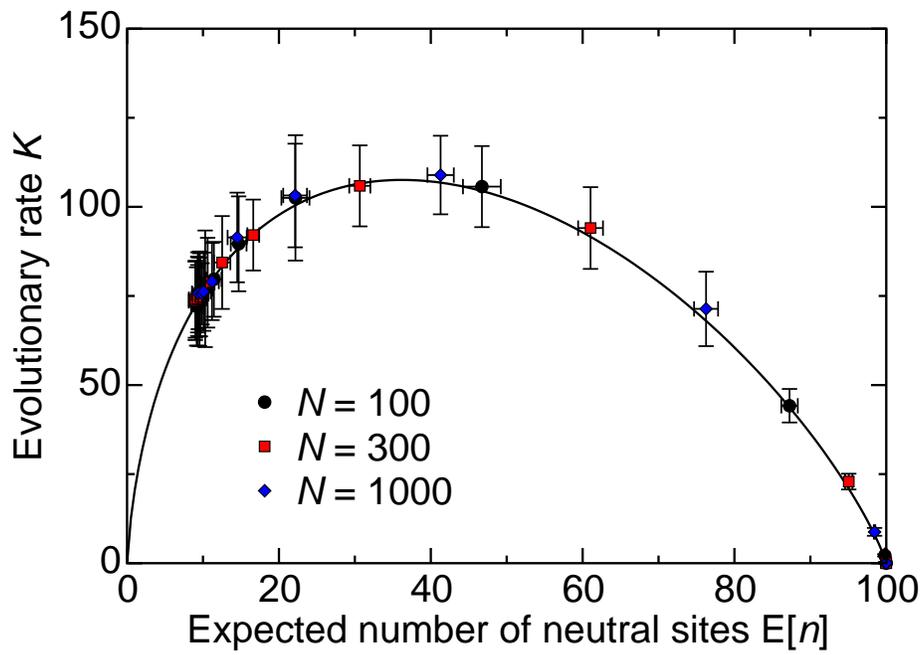}}
\caption{\label{fig:rate-vs-neutr}Evolutionary rate versus expected number of neutral sites $\expect{n}$, for population sizes $N=100, 300, 1000$ ($\alpha=0.1$). The solid line stems from Eq.~\eqref{eq:K-of-n} in the Appendix. Error bars indicate standard errors.
}
\end{figure}

Finally, the calculation in the Appendix predicts that the evolutionary rate should be independent of $N$ if we plot it as a function of the expected number of neutral sites $\expect{n}$ rather than a function of $A$. Figure~\ref{fig:rate-vs-neutr} shows that this prediction is indeed accurate. We find that there are two distinct regimes for the evolutionary rate. For small $\expect{n}$, the evolutionary rate increases with $\expect{n}$. This increase is caused by the increased availability of neutral mutations with growing $\expect{n}$. However, even though we can calculate what the evolutionary rate would be for arbitrarily small $\expect{n}$, in equilibrium $\expect{n}$ will never be below its limiting value for small $A$, $\alpha L/(\alpha+1)$. For large $\expect{n}$, the behavior of $K$ is reversed, and now it decreases with increasing $\expect{n}$. The decay comes about because in this regime, even though there are many mutations which do not disrupt the fold of a properly translated protein, these mutations increase the amount of mistranslated, misfolded proteins, and thus are selected against. The quantity $\expect{n}$ can get arbitrarily close to $L$ (or $l$ for $m>0$), and therefore $K$ can decay to almost zero if $A$ is sufficiently large.

Throughout this study, we found good agreement among the numerical simulations, Sella-Hirsh theory, and our simple analytical approximation. Some discrepancies appeared between theory and simulations for the largest population size ($N=1000$) and for very small $\alpha$ in conjunction with large $A$. We attribute the former to a violation of the condition $NU\ll 1$, which must be satisfied for both Sella-Hirsh theory and our approximate solution. We carried out our simulations with a mutation rate of $U=0.001$, which means that $NU=1$ for $N=1000$. The latter discrepancies are caused by insufficient equilibration time. For large $A$, the number of neutral sites $n$ always approaches $L$, irrespective of population size or $\alpha$. However, the smaller $\alpha$ is, the lower the probability that a mutation occurs which increases $n$. Therefore, the equlibration time needed at large $A$ grows without bound as $\alpha$ decreases. We did additional simulations in this regime, and found that the simulation results approached the predicted quantities with increasing equilibration time (data not shown).
\bigskip

\centerline{DISCUSSION}
\bigskip

We have developed a simple model that describes the slowdown of the rate of evolution of highly expressed genes under selective pressure for translational robustness. We have studied the model with numerical simulations and have solved the model exactly using Sella-Hirsh theory. We have also developed a simple analytic approximation that is in excellent agreement with the predictions from Sella-Hirsh theory, and is valid for the entire range of possible parameter values (as long as $NU \ll 1$).

The model abstracts a previous thermodynamic model of protein mutational tolerance introduced by \citet{Bloometal2005} in which mutations may leave unperturbed or destabilize the protein's native structure (common) or stabilize it (uncommon).  Increases in stability tend to increase the number of sites at which substitutions can be tolerated, so-called neutral sites, while decreases in stability usually cause misfolding or decrease the number of neutral sites.  In the present work, we have modeled neutral sites directly.  In doing so, we only allow stepwise changes in neutral sites, sacrificing treatment of large stability changes that might radically alter the number of neutral sites and the potential stability dependence of mutational effects in order to gain analytical tractability.

Our results show a clear example of the paradox cited in the Introduction (Fig.~\ref{fig:rate-vs-neutr}): Given selection against the costs of protein misfolding, genes simultaneously become more mutationally tolerant (larger number of neutral sites, $n$) but evolve slower as if fewer mutations were tolerated.  The paradox's resolution requires disentangling two kinds of mutational tolerance, one of which captures the likelihood of loss of protein function due to a mutation while the other quantifies the fitness cost of that mutation, the cost which ultimately determines evolutionary rate.  When protein misfolding imposes minimal fitness costs, as is the case with low-expression proteins, the proportion of mutations which preserve protein function govern the rate of evolution (Fig.~\ref{fig:rate-vs-neutr}, left).  However, at high expression levels, fitness costs of mutations which preserve protein function grow large and can become the dominant determinant of evolutionary rate (Fig.~\ref{fig:rate-vs-neutr}, right).

This observation has an important corollary.  When selection for translational robustness is weak, functional loss is likely the main determinant of fitness costs.  Thus, our results suggest that evolutionary conservation of sites in low-expression proteins may be more likely to indicate functional importance than similar conservation at sites in high-expression proteins.

Our simple model produces an exponential decline in evolutionary rate with increasing expression level, whereas in yeast, a power law better describes this relationship \citep{Drummondetal2005}.  Several possibilities may explain the discrepancy. First, our model assumes a symmetric binomial distribution of the number of neutral sites, but the distribution for real proteins may be skewed or heavy-tailed. Second, the cost of additional misfolded proteins may not be independent of the number of already misfolded proteins. For example, misfolded proteins form toxic aggregates \citep{Bucciantinietal2002}, and aggregation is not a linear function of protein concentration.  Finally, differences in protein structure and function between high- and low-expression proteins may play a role.  \citet{Drummondetal2005} have previously examined differences between functionally and structurally similar paralogs and found a similar power-law relationship.  However, more subtle but important differences may separate paralogs and influence their evolution.

Our results here demonstrate that profound differences in protein evolutionary rates can arise even in the absence of functional and structural differences and when variables such as protein length, the
translation error rate, and the underlying distribution of the number of neutral sites are held constant.  In real genomes, of course, all these features vary and some, perhaps all, are under selection.  The value of the model is its utility in explaining why highly expressed proteins evolve slowly across taxa \citep{Drummondetal2005}.

Interestingly, our model reveals two evolutionary-rate regimes (Figs.~\ref{fig:Nvar} and \ref{fig:alphavar}), one in which rates remain relatively constant with increasing protein production, and another in which rates decline precipitously.  In yeast, virtually all genes appear to fall in the latter regime, as the evolutionary rate declines almost immediately from low to high expression \citep{Drummondetal2005}, raising the possibility of genome-level selection in this direction.  If yeast protein synthesis levels reflect organismal needs and cannot be freely modulated, as seems likely, and protein synthesis costs dominate cellular energy consumption, as evidence suggests \citep{Princiottaetal2003}, the remaining genome-level target for selection is on the fitness cost per misfolded protein, $c$.  Decreasing $c$ pushes genes away from the decline to where cost differences become negligible, whereas increasing $c$ pushes genes toward the decline, amplifying the cost difference between high- and low-expression proteins.  One way to decrease $c$ is to drive translation errors down to negligible levels.  Another is to maintain a quality-control apparatus (e.g.\ chaperones and proteases) with so much excess bandwidth that cost differences associated with variability in protein misfolding become negligible.  Evidence suggests that both strategies for decreasing $c$ impose significant fitness penalties. Decreasing translational error rates can be easily accomplished, often with a single ribosomal mutation \citep{Alksneetal93}, but ribosomal accuracy and growth rate are often negatively correlated, presumably through the intrinsic speed/accuracy tradeoff inherent in ribosomal proofreading \citep{Kurland92}.  Maintenance of a chaperone fleet large enough to dilute out misfolding cost differences would divert enormous cellular resources for little benefit, and the massive induction of chaperones after heat shock suggests that cellular chaperone levels do not have much remaining bandwidth under normal conditions.  Overall, it seems plausible that the steep decline observed in yeast's evolutionary-rate--expression relationship reflects a balance favoring a relatively high cost per misfolded protein $c$.  Costly translational accuracy and quality control machinery may be reduced so long as the increased errors and reduced folding assistance can be compensated for; translational robustness provides that compensation---essentially for free---but is ultimately limited by mutation pressure away from robust sequences and by the fundamental intolerance of proteins to at least some errors.

Our model distinguishes between the number of polypeptides produced per gene, $x$, and the abundance of functional proteins, $A$, yet our approximate solution essentially equates these quantities with only minor accuracy loss.  The approximation works for two reasons.  First, misfolded polypeptides impose a negligible cost for low-abundance proteins, while for high-abundance proteins, misfolded polypeptides are rare because of selection for translational robustness.  We expect these nontrivial results to hold for many organisms.  Second, in our model, the number of translation events $x$, the primary determinant of the number of mistranslated proteins, is estimated accurately by $A$, a situation unlikely to hold for most organisms. Protein abundance reflects a balance of ongoing translation and turnover \citep{Greenbaumetal2003}, such that a high abundance can result from either moderate translational frequency and long protein half-life or from rapid translation and short half-life.  Because half-lives can vary over orders of magnitude \citep{HargroveSchmidt89}, abundance and translation frequency may only weakly correlate in real organisms.  Among protein abundance, mRNA expression level, and translation frequency, we hypothesize that the latter, even though difficult to measure, will best predict evolutionary rate.

\citet{Burgeretal2005} recently studied a question closely related to the present paper, asking why phenotypic mutation rates (corresponding to the translational error rate in the present paper) are much higher than genotypic mutation rates. Within the framework of their model, \citet{Burgeretal2005} found very little pressure for reduction of phenotypic mutation rates below a certain threshold. Even though we keep the translational error rate constant in our model, we can consider a change in the number of neutral sites $n$ as a change in the phenotypic mutation rate, and thus compare our results to those of \citet{Burgeretal2005}. In contrast to their conclusions, we find that the fraction of neutral sites, $n/L$, quickly rises to the maximum possible for highly expressed genes, thus reducing the phenotypic mutation rate to zero except when some sites cannot be made neutral. We believe that the differences in results are caused by differences in the way in which we and \citet{Burgeretal2005} treated costs of erroneously translated proteins in our models. \citet{Burgeretal2005} consider the total cost of protein synthesis, but do not consider additional penalties imposed by misfolded proteins, not only for their recognition and cleanup by the quality-control system but also for their innate toxicity \citep{Bucciantinietal2002}. Clearly, if we neglect these unique costs, then the only pressure to reduce the phenotypic mutation rate is to reduce the cost of synthesis for misfolded proteins, and this pressure will be weak if this cost is only a small proportion of the total cost of protein synthesis. In our model, on the other hand, we have focused exclusively on costs of misfolded proteins apart from their synthesis costs, implicitly assuming that the total cost of protein synthesis is approximately equal to the cost of synthesis of functional proteins, and that the benefit of the functional proteins will pay for their synthesis. We believe that there is indeed a strong selective pressure to reduce the phenotypic mutation rate for highly expressed genes, but that it is cheaper for cells to evolve translationally robust genes than to evolve highly accurate transcription and translation machinery.

Can translational robustness really be obtained cheaply?  \citet{Drummondetal2005} have suggested that increased protein stability both confers mutational tolerance and constrains sequence evolution.  
Increasing protein stability, that is, decreasing the free energy of folding $\Delta G_f$, provides a plausible mechanism for obtaining translational robustness for numerous reasons.  First, increased stability leads to increased mutational tolerance and can be obtained by point mutations \citep{Bloometal2005}.  Second, the stability-increase mechanism is sufficiently general to encompass proteins of diverse functions and to operate in a wide range of organisms.  Third, stability is free in the sense that obtaining a protein with lower $\Delta G_f$ requires only a chance mutation.  While many researchers have noted an apparent tradeoff between protein stability and enzymatic activity, it is crucial to emphasize that this trend may be statistical rather than intrinsic: Because both high activity and high stability are rare properties, mutations that improve both are exceedingly unlikely unless both are constrained \citep{Giveretal1998}.  Selection for translational robustness provides precisely that dual constraint, and because many millions of mutations may be screened over evolutionary time, the very few resulting in highly expressed proteins with increased stability (conferring tolerance to translation errors) and uncompromised activity will be found.  Finally, the very rarity of such stabilizing mutations provides a measure of the scarcity of highly stable proteins available for exploration by evolutionary drift.  If increased stability is a dominant response to the need for mutational tolerance in highly expressed proteins, it will restrict drift and slow evolution relative to less-constrained low-expression proteins.

\bigskip

\centerline{ACKNOWLEDGMENTS}
\bigskip
C.O.W.~was supported by NIH grant AI 065960 and D.A.D.~was supported by NIH National Research Service Award 5 T32 MH19138.  D.A.D. acknowledges, with gratitude, the support of Frances Arnold.


\begin{thebibliography}{35}
\providecommand{\natexlab}[1]{#1}
\providecommand{\url}[1]{\texttt{#1}}
\providecommand{\urlprefix}{URL }
\expandafter\ifx\csname urlstyle\endcsname\relax
  \providecommand{\doi}[1]{doi:\discretionary{}{}{}#1}\else
  \providecommand{\doi}{doi:\discretionary{}{}{}\begingroup
  \urlstyle{rm}\Url}\fi
\providecommand{\eprint}[2][]{\url{#2}}

\bibitem[{\textsc{Agrafioti} \emph{et~al.}(2005)\textsc{Agrafioti},
  \textsc{Swire}, \textsc{Abbott}, \textsc{Huntley}, \textsc{Butcher} and
  \textsc{Stumpf}}]{Agrafiotietal2005}
\textsc{Agrafioti, I.}, \textsc{J.~Swire}, \textsc{J.~Abbott},
  \textsc{D.~Huntley}, \textsc{S.~Butcher} and \textsc{M.~P.~H. Stumpf}, 2005
  Comparative analysis of the \emph{Saccharomyces cerevisiae} and
  \emph{Caenorhabditis elegans} protein interaction networks.
\newblock BMC Evol. Biol. \textbf{5:} 23.

\bibitem[{\textsc{Akashi}(1994)}]{Akashi94}
\textsc{Akashi, H.}, 1994 Synonymous codon usage in \emph{Drosophila
  melanogaster}: {Natural} selection and translational accuracy.
\newblock Genetics \textbf{136:} 927-935.

\bibitem[{\textsc{Akashi}(2001)}]{Akashi2001}
\textsc{Akashi, H.}, 2001 Gene expression and molecular evolution.
\newblock Current Opinion in Genetics \& Development \textbf{11:} 660-666.

\bibitem[{\textsc{Alksne} \emph{et~al.}(1993)\textsc{Alksne}, \textsc{Anthony},
  \textsc{Liebman} and \textsc{Warner}}]{Alksneetal93}
\textsc{Alksne, L.~E.}, \textsc{R.~A. Anthony}, \textsc{S.~W. Liebman} and
  \textsc{J.~R. Warner}, 1993 An accuracy center in the ribosome conserved over
  2 billion years.
\newblock Proc. Natl. Acad. Sci. USA \textbf{90:} 9538-9541.

\bibitem[{\textsc{Bloom} and \textsc{Adami}(2003)}]{BloomAdami2003}
\textsc{Bloom, J.~D.} and \textsc{C.~Adami}, 2003 Apparent dependence of
  protein evolutionary rate on number of interactions is linked to biases in
  protein-protein interactions data sets.
\newblock BMC Evol. Biol. \textbf{3:} 21.

\bibitem[{\textsc{Bloom} \emph{et~al.}(2005)\textsc{Bloom}, \textsc{Silberg},
  \textsc{Wilke}, \textsc{Drummond}, \textsc{Adami} and
  \textsc{Arnold}}]{Bloometal2005}
\textsc{Bloom, J.~D.}, \textsc{J.~J. Silberg}, \textsc{C.~O. Wilke},
  \textsc{D.~A. Drummond}, \textsc{C.~Adami} and \textsc{F.~H. Arnold}, 2005
  Thermodynamic prediction of protein neutrality.
\newblock Proc. Natl. Acad. Sci. USA \textbf{102:} 606-611.

\bibitem[{\textsc{Bucciantini} \emph{et~al.}(2002)\textsc{Bucciantini},
  \textsc{Giannoni}, \textsc{Chiti}, \textsc{Baroni}, \textsc{Formigli},
  \textsc{Zurdo}, \textsc{Taddei}, \textsc{Ramponi}, \textsc{Dobson} and
  \textsc{Stefani}}]{Bucciantinietal2002}
\textsc{Bucciantini, M.}, \textsc{E.~Giannoni}, \textsc{F.~Chiti},
  \textsc{F.~Baroni}, \textsc{L.~Formigli}, \textsc{J.~Zurdo},
  \textsc{N.~Taddei}, \textsc{G.~Ramponi}, \textsc{C.~M. Dobson} and
  \textsc{M.~Stefani}, 2002 Inherent toxicity of aggregates implies a common
  mechanism for protein misfolding diseases.
\newblock Nature \textbf{416:} 507-511.

\bibitem[{\textsc{B\"urger} \emph{et~al.}(2005)\textsc{B\"urger},
  \textsc{Willensdorfer} and \textsc{Nowak}}]{Burgeretal2005}
\textsc{B\"urger, R.}, \textsc{M.~Willensdorfer} and \textsc{M.~A. Nowak}, 2005
  Why are phenotypic mutation rates much higher than genotypic mutation rates?
\newblock Genetics, in press. \doi{10.1534/genetics.105.046599}.

\bibitem[{\textsc{Drummond} \emph{et~al.}(2005)\textsc{Drummond},
  \textsc{Bloom}, \textsc{Adami}, \textsc{Wilke} and
  \textsc{Arnold}}]{Drummondetal2005}
\textsc{Drummond, D.~A.}, \textsc{J.~D. Bloom}, \textsc{C.~Adami},
  \textsc{C.~O. Wilke} and \textsc{F.~H. Arnold}, 2005 Why highly
  expressed proteins evolve slowly.
\newblock Proc. Natl. Acad. Sci. USA \textbf{102:} 14338-14343.

\bibitem[{\textsc{Drummond} \emph{et~al.}(2006)\textsc{Drummond},
  \textsc{Raval} and \textsc{Wilke}}]{Drummondetal2005b}
\textsc{Drummond, D.~A.}, \textsc{A.~Raval} and \textsc{C.~O. Wilke},
  2006 A single determinant dominates the rate of yeast protein
  evolution. Mol. Biol. Evol., in press.

\bibitem[{\textsc{Fraser} \emph{et~al.}(2002)\textsc{Fraser}, \textsc{Hirsh},
  \textsc{Steinmetz}, \textsc{Scharfe} and \textsc{Feldman}}]{Fraseretal2002}
\textsc{Fraser, H.~B.}, \textsc{A.~E. Hirsh}, \textsc{L.~M. Steinmetz},
  \textsc{C.~Scharfe} and \textsc{M.~W. Feldman}, 2002 Evolutionary rate in the
  protein interaction network.
\newblock Science \textbf{296:} 750-752.

\bibitem[{\textsc{Giver} \emph{et~al.}(1998)\textsc{Giver},
  \textsc{Gershenson}, \textsc{Freskgard} and \textsc{Arnold}}]{Giveretal1998}
\textsc{Giver, L.}, \textsc{A.~Gershenson}, \textsc{P.-O. Freskgard} and
  \textsc{F.~H. Arnold}, 1998 Directed evolution of a thermostable esterase.
\newblock Proc. Natl. Acad. Sci. USA \textbf{95:} 12809-12813.

\bibitem[{\textsc{Greenbaum} \emph{et~al.}(2003)\textsc{Greenbaum},
  \textsc{Colangelo}, \textsc{Williams} and
  \textsc{Gerstein}}]{Greenbaumetal2003}
\textsc{Greenbaum, D.}, \textsc{C.~Colangelo}, \textsc{K.~Williams} and
  \textsc{M.~Gerstein}, 2003 Comparing protein abundance and {mRNA} expression
  levels on a genomic scale.
\newblock Genome Biol. \textbf{4:} 117.

\bibitem[{\textsc{Hahn} \emph{et~al.}(2004)\textsc{Hahn}, \textsc{Conant} and
  \textsc{Wagner}}]{Hahnetal2004}
\textsc{Hahn, M.~W.}, \textsc{G.~C. Conant} and \textsc{A.~Wagner}, 2004
  Molecular evolution in large genetic networks: Does connectivity equal
  constraint?
\newblock J. Mol. Evol. \textbf{58:} 203-211.

\bibitem[{\textsc{Hahn} and \textsc{Kern}(2005)}]{HahnKern2005}
\textsc{Hahn, M.~W.} and \textsc{A.~D. Kern}, 2005 Comparative genomics of
  centrality and essentiality in three eukaryotic protein-interaction networks.
\newblock Mol. Biol. Evol. \textbf{22:} 803-806.

\bibitem[{\textsc{Hargrove} and \textsc{Schmidt}(1989)}]{HargroveSchmidt89}
\textsc{Hargrove, J.~L.} and \textsc{F.~H. Schmidt}, 1989 The role of {mRNA}
  and protein stability in gene expression.
\newblock FASEB J. \textbf{3:} 2360-2370.

\bibitem[{\textsc{Hirsh} and \textsc{Fraser}(2001)}]{HirshFraser2001}
\textsc{Hirsh, A.~E.} and \textsc{H.~B. Fraser}, 2001 Protein dispensability
  and rate of evolution.
\newblock Nature \textbf{411:} 1046-1049.

\bibitem[{\textsc{Hurst} and \textsc{Smith}(1999)}]{HurstSmith99}
\textsc{Hurst, L.~D.} and \textsc{N.~G.~C. Smith}, 1999 Do essential genes
  evolve slowly?
\newblock Curr. Biol. \textbf{9:} 747-750.

\bibitem[{\textsc{Jordan} \emph{et~al.}(2002)\textsc{Jordan}, \textsc{Rogozin},
  \textsc{Wolf} and \textsc{Koonin}}]{Jordanetal2002}
\textsc{Jordan, I.~K.}, \textsc{I.~B. Rogozin}, \textsc{Y.~I. Wolf} and
  \textsc{E.~V. Koonin}, 2002 Essential genes are more evolutionarily conserved
  than are nonessential genes in bacteria.
\newblock Genome Res. \textbf{12:} 962-968.

\bibitem[{\textsc{Jordan} \emph{et~al.}(2003)\textsc{Jordan}, \textsc{Wolf} and
  \textsc{Koonin}}]{Jordanetal2003}
\textsc{Jordan, I.~K.}, \textsc{Y.~I. Wolf} and \textsc{E.~V. Koonin}, 2003 No
  simple dependence between protein evolution rate and the number of
  protein-protein interactions: only the most prolific interactors tend to
  evolve slowly.
\newblock BMC Evol. Biol. \textbf{3:} 1.

\bibitem[{\textsc{Kellogg} and \textsc{Juliano}(1997)}]{KelloggJuliano1997}
\textsc{Kellogg, E.} and \textsc{N.~Juliano}, 1997 The structure and function of RuBisCO and
their implications for systematic studies.
\newblock Am. J. Botany \textbf{84:} 413-428.

\bibitem[{\textsc{Kimura}(1983)}]{Kimura83}
\textsc{Kimura, M.}, 1983 The neutral theory of molecular evolution.
\newblock Cambridge University Press.

\bibitem[{\textsc{Kurland}(1992)}]{Kurland92}
\textsc{Kurland, C.~G.}, 1992 Translational accuracy and the fitness of
  bacteria.
\newblock Annu Rev Genet \textbf{26:} 29-50.

\bibitem[{\textsc{Marais} and \textsc{Duret}(2001)}]{MaraisDuret2001}
\textsc{Marais, G.} and \textsc{L.~Duret}, 2001 Synonymous codon usage,
  accuracy of translation, and gene length in \emph{Caenorhabditis elegans}.
\newblock J. Mol. Evol. \textbf{52:} 275-280.

\bibitem[{\textsc{Ohta}(1992)}]{Ohta92}
\textsc{Ohta, T.}, 1992 The nearly neutral theory of molecular evolution.
\newblock Annu. Rev. Ecol. Syst. \textbf{23:} 263-286.

\bibitem[{\textsc{Pal} \emph{et~al.}(2001)\textsc{Pal}, \textsc{Papp} and
  \textsc{Hurst}}]{Paletal2001}
\textsc{Pal, C.}, \textsc{B.~Papp} and \textsc{L.~D. Hurst}, 2001 Highly
  expressed genes in yeast evolve slowly.
\newblock Genetics \textbf{158:} 927-931.

\bibitem[{\textsc{Pal} \emph{et~al.}(2003)\textsc{Pal}, \textsc{Papp} and
  \textsc{Hurst}}]{Paletal2003}
\textsc{Pal, C.}, \textsc{B.~Papp} and \textsc{L.~D. Hurst}, 2003 Rate of
  evolution and gene dispensability.
\newblock Nature \textbf{421:} 496-497.

\bibitem[{\textsc{Precup} and \textsc{Parker}(1987)}]{PrecupParker87}
\textsc{Precup, J.} and \textsc{J.~Parker}, 1987 Missense misreading of
  asparagine codons as a function of codon identity and context.
\newblock J. Biol. Chem. \textbf{262:} 11351-11355.

\bibitem[{\textsc{Princiotta} \emph{et~al.}(2003)\textsc{Princiotta},
  \textsc{Finzi}, \textsc{Qian}, \textsc{Gibbs}, \textsc{Schuchmann},
  \textsc{Buttgereit}, \textsc{Bennink} and
  \textsc{Yewdell}}]{Princiottaetal2003}
\textsc{Princiotta, M.~F.}, \textsc{D.~Finzi}, \textsc{S.~B. Qian},
  \textsc{J.~Gibbs}, \textsc{S.~Schuchmann}, \textsc{F.~Buttgereit},
  \textsc{J.~R. Bennink} and \textsc{J.~W. Yewdell}, 2003 Quantitating protein
  synthesis, degradation, and endogenous antigen processing.
\newblock Immunity \textbf{18:} 343-354.

\bibitem[{\textsc{Rocha } and \textsc{Danchin}(2004)}]{RochaDanchin2004}
\textsc{Rocha , E. P.~C.} and \textsc{A.~Danchin}, 2004 An analysis of
  determinants of amino acids substitution rates in bacterial proteins.
\newblock Mol. Biol. Evol. \textbf{21:} 108-116.

\bibitem[{\textsc{Sella} and \textsc{Hirsh}(2005)}]{SellaHirsh2005}
\textsc{Sella, G.} and \textsc{A.~E. Hirsh}, 2005 The application of
  statistical physics to evolutionary biology.
\newblock Proc. Natl. Acad. Sci. USA \textbf{102:} 9541-9546.

\bibitem[{\textsc{Spreitzer}(1993)}]{Spreitzer93}
\textsc{Spreitzer, R.~J.}, 1993 Genetic dissection of {Rubisco} structure and
  function.
\newblock Annu. Rev. Plant Physiol. Plant Mol. Biol. \textbf{44:} 411-434.

\bibitem[{\textsc{Wall} \emph{et~al.}(2005)\textsc{Wall}, \textsc{Hirsh},
  \textsc{Fraser}, \textsc{Kumm}, \textsc{Giaever}, \textsc{Eisen} and
  \textsc{Feldman}}]{Walletal2005}
\textsc{Wall, D.~P.}, \textsc{A.~E. Hirsh}, \textsc{H.~B. Fraser},
  \textsc{J.~Kumm}, \textsc{G.~Giaever}, \textsc{M.~B. Eisen} and \textsc{M.~W.
  Feldman}, 2005 Functional genomic analysis of the rates of protein evolution.
\newblock Proc. Natl. Acad. Sci. USA \textbf{102:} 5483-5488.

\bibitem[{\textsc{Wilke}(2004)}]{Wilke2004}
\textsc{Wilke, C.~O.}, 2004 Molecular clock in neutral protein evolution.
\newblock BMC Genetics \textbf{5:} 25.

\bibitem[{\textsc{Wilke} \emph{et~al.}(2005)\textsc{Wilke}, \textsc{Bloom},
  \textsc{Drummond} and \textsc{Raval}}]{Wilkeetal2005}
\textsc{Wilke, C.~O.}, \textsc{J.~D. Bloom}, \textsc{D.~A. Drummond} and
  \textsc{A.~Raval}, 2005 Predicting the tolerance of proteins to random amino
  acid substitution.
\newblock Biophys. J. \textbf{89:} 3714-3720.

\bibitem[{\textsc{Zhang} and \textsc{He}(2005)}]{ZhangHe2005}
\textsc{Zhang, J.} and \textsc{X.~He}, 2005 Significant impact of protein
  dispensability on the instantaneous rate of protein evolution.
\newblock Mol. Biol. Evol. \textbf{22:} 1147-1155.

\end{thebibliography}

\begin{appendix}
\centerline{APPENDIX}
\bigskip

Here we derive an expression for the evolutionary rate $K$ as a function of the number of neutral sites $n*$ rather than the protein abundance $A$. For the remainder of this appendix, we drop the superscript from $n^*$ for simplicity. We begin by noting that Eq.~\eqref{eq:n-approx} implies
\begin{equation}\label{eq:pim-of-pip}
  n\pi_- = \alpha(L-n)\pi_+\,.
\end{equation}
Further, note that we can write, for sufficiently large $N$,
\begin{equation}
  \pi_- = e^{-2NA\tau}\pi_+\,.
\end{equation}
Therefore, $e^{-2NA\tau}=\alpha(L-n)/n$. We can solve this expression for $A$ and find
\begin{equation}\label{eq:A-of-n}
  A=\frac{-1}{2N\tau}\ln[\alpha(L-n)/n]\,.
\end{equation}
After inserting Eq.~\eqref{eq:A-of-n} into the definition of $\pi_-$, we obtain
\begin{equation}
  \pi_- = \frac{1-\exp\Big(\frac{-1}{N}\ln[\alpha(L-n)/n]\Big)} 
    {1-\exp\Big(-\ln[\alpha(L-n)/n]\Big)} \,.
\end{equation}
We expand this expression for large $N$, replacing the exponential in the numerator by the first two terms of its Taylor series, and find
\begin{equation}\label{eq:pim-final}
  \pi_- = \frac{1}{N}\frac{\alpha(L-n)}{\alpha L-(\alpha+1)n} \ln[\alpha(L-n)/n]\,.
\end{equation}
Now, after inserting Eqs.~\eqref{eq:pim-of-pip} and~\eqref{eq:pim-final} into Eq.~\eqref{eq:evol-rate}, we obtain for the expected evolutionary rate
\begin{equation}\label{eq:K-of-n}
  K = 2U\frac{\alpha n (L-n)}{\alpha L^2-(\alpha+1)Ln}\ln[\alpha(L-n)/n]\,.
\end{equation}
Note that this expression is independent of the population size $N$. Even though we have derived it under the assumption that $N$ is large, we find that it works very well even for moderate population sizes of 100 or less.

\end{appendix}

\cleardoublepage

\begin{table}
\begin{center}
\begin{tabular}{|c|l|}
\hline
$\alpha$ & probability of turning a non-neutral into a neutral site \\
\hline
$A$ & number (abundance) of folded proteins \\
\hline
$f$ & fraction of folded proteins \\
\hline
$F$ & Boltzmann factor \\
\hline
$K$ & number of amino-acid substitutions since most\\[-.1in] &recent common ancestor (evolutionary rate) \\
\hline
$l$ & number of possibly-neutral sites \\
\hline
$L$ & protein length \\
\hline
$m$ & number of never-neutral sites \\
\hline
$n$ & number of neutral sites \\
\hline
$N$ & population size \\
\hline
$t$ & time, in generations \\
\hline
$\tau$ & translation error probability per site \\
\hline
$U$ & mutation rate per sequence \\
\hline
$w_n$ & fitness of sequence with $n$ neutral sites\\
\hline
\end{tabular}
\end{center}
\caption{Variables used in this work.}\label{tab:variables}
\end{table}

\end{document}